\documentstyle[aps,eqsecnum,pra,epsf]{revtex} \twocolumn

\input{epsf}

\newcommand{\lkl}{\left(}
\newcommand{\rkl}{\right)}

\newcommand{\omegarf}{\omega_{\rm{rf}}}
\newcommand{\omegatrap}{\omega_{\rm{trap}}}

\begin{document}

\title{Output from an atom laser: theory vs. experiment}
\author{Jens Schneider and Axel Schenzle}
\address{Max-Planck-Institut f\"ur Quantenoptik,
  Hans-Kopfermann-Stra\ss{}e 1, D-85748 Garching, Germany and \\
    Sektion Physik, Universit\"at M\"unchen, Theresienstra\ss{}e 37, D-80333
  M\"unchen, Germany \\ (Fax: +49-89/32905-200,
  E-mail: jens.schneider@mpq.mpg.de)}

\date{submitted to App. Phys. B}

\maketitle

\begin{abstract}
  Atom lasers based on rf-outcoupling can be described by a set of
  coupled generalized Gross-Pitaevskii equations (GPE). We compare the
  theoretical predictions obtained by numerically integrating the
  time-depen\-dent GPE of an effective one-dimensional model with
  recently measured experimental data for the F=2 and F=1 states of
  Rb-87. We conclude that the output of a rf-atom laser can be well
  described by this model.
\end{abstract}

\narrowtext

The discovery of Bose-Einstein condensation (BEC) in trapped atomic
gases \cite{ANDE95,BRAD95,DAVIS95} and the subsequent demonstration of
macroscopic coherence of Bose condensed gases \cite{ANDR97:1,ROEH97}
have raised a lot of interest in creating coherent atomic beams
using such condensates. Experimental setups based on Bose-Einstein
condensates that produce coherent atomic beams are now called {\em
  atom lasers}. The first experimental realization \cite{MEW97} was
based on a pulsed outcoupler; strong radio-frequency pulses were used
to flip magnetically trapped atoms to untrapped states such that they
could leak out of the trap. The method of rf-outcoupling has been
extensively investigated theoretically
\cite{BALL97,HOPE97,NARA97:2,STEC98:1,ZHAN98:1,ZHAN98:2,GRAH99:1},
whereas the possibility of using Raman transitions \cite{MOY97} has
not been treated in such detail. Most recently, the outcoupling process
was treated taking also thermal excitations within the trap into
account \cite{JAPHA99:1}.

Experiments have also made some re\-mark\-able prog\-ress. In
\cite{ANDE98:1} a Bose condensate is released from a magnetic trap
into an optical standing wave leading to mode-locked like pulses of
atoms falling downwards. A method using Raman pulses is demonstrated
in \cite{HAGL99:1}. There, the outcoupled atoms are moving
horizontally and are pulsed with a very high frequency resulting in a
quasi-continuous beam. Finally, Bloch {\em et al.} \cite{BLOCH99:1} use
a rf-outcoupler within a very stable magnetic setup, allowing to
operate the atom laser in a regime close to the weak-coupling regime
dealt with in \cite{STEC98:1}. As in all other cases there is no refill
mechanism for the Bose condensate in the trap that balances the
outcoupling loss. Nevertheless, the outcoupling rate is so low that
the output is similar to a cw-laser with a slowly decreasing intensity.

In the following, we want to model the situation in
\cite{BLOCH99:1}. The trap is elongated horizontally. As the atomic
beam is directed downward, we approximate the trap with a 1D model in
the vertical $z$-direction taking account of the other two dimensions in a 
way explained in the next section.

\section{Theory of coupled Gross-Pitaevskii equations}
\label{sec:theory}

Condensates of weakly interacting Bose gases at zero temperature are
well described by the Gross-Pitaevskii equation (see \cite{DALV98:1}
and references therein). As it is a nonlinear Schr\"odinger equation
for a macroscopic wavefunction it not only accounts for the density
properties but also for the coherence inherent in Bose condensates. In
\cite{BALL97} a generalization of the GPE was introduced to treat a
system of trapped and untrapped states of a Bose gas coupled via an rf
field with a frequency $\omegarf$. Here, we want to model an atom
laser for Rb-87 in its $F=2$ or $F=1$ hyperfine-manifold. The $2F+1$
Zeeman sublevels are represented by macroscopic wavefunctions $\psi_m$
($m\in\{-F,\ldots,F\}$) so that the system of generalized GPEs after
applying the transformation $\psi_m(t) \rightarrow e^{-i m \omegarf t}
\psi_m(t)$ and the rotating wave approximation reads
\begin{eqnarray}
  \label{eq:gpe3d}
  i \hbar \frac{\partial}{\partial t} \psi_m(\vec r, t) &=&
  \Bigg( -\frac{\hbar^2\nabla^2}{2M} + V_m(\vec r) \nonumber \\
  && \hspace{0.35cm} + \hbar m \omegarf
     + U ||\psi(\vec r, t)||^2 \Bigg) \psi_m(\vec r, t) \nonumber \\
  && \hspace{-1.4cm} + \hbar \Omega
  \sum_{m'} \lkl c_{m'} \delta_{m,m'+1} + c_m \delta_{m,m'-1} \rkl
  \psi_{m'} (\vec r, t)\;.
\end{eqnarray}
Thereby we have assumed that all Zeemann levels interact with the same
s-wave scattering length $a_0 = 110\, a_{\rm Bohr}$ so that $U =
4\pi\hbar^2 a_0 N/M$. The total density devided by the number $N$ of atoms in
the system is denoted by $||\psi(\vec r, t)||^2 = \sum_m |\psi_m(\vec
  r, t)|^2$ (we normalize to 1, not to N). The Rabi frequency is
determined by the strength $B_{\rm rf}$ of the rf-field and given
by $\hbar \Omega = g_F \mu_{\rm Bohr} B_{\rm rf}/2$. It is modified by
the matrix elements of the angular operators $F_{\pm}$ represented by
$c_m = \sqrt{F(F+1)-m(m+1)}$. For $F=1$, this results in $c_m =
\sqrt{2}$; for $F=2$, $c_m$ takes values 2 and $\sqrt{6}$ depending
on $m$ so the Rabi frequencies differ for different $m\rightarrow m'$
transitions. 

Ioffe-type traps often used in experiments are usually elongated in
the horizontal plane being axial symetric in the remaining directions.
As mentioned above, we want to model an atom laser based on such traps
by 1D GPEs where the axis under consideration is the vertical axis of
the trap. We choose coordinates where the $z$-axis points {\em
  downwards}, the long axis of the trap is denoted by $y$, the short
horizontal one by $x$. Taking gravity into account, the total
effective potentials in $z$-direction then are given by

\begin{eqnarray}
  V_{m,{\rm eff}}(z,t) &=& {\rm sgn}(g_F)\, m\,M \omega_z^2 z^2/2
  + m \hbar \Delta \nonumber\\
                       && \hspace{2cm} - Mgz + g_{\rm 1D} ||\psi(z,t)||^2\,.
\end{eqnarray}
$\Delta= \hbar\omegarf - V_{\rm off}$ is simply the detuning of the
transitions
($V_{\rm off} = -g_F \mu_{\rm Bohr} B_{\rm off}$, $\omegarf > 0$
for $F=1$ and $\omegarf<0$ for $F=2$).

The effective interaction strength $g_{\rm 1D}$ is determined in such
a way that the chemical potential in Thomas-Fermi approximation in
the 1D model equals that for the full 3D situation for $N$ particles
in a trapped state $m_{\rm trap}$.  It turns out to be equal to
\begin{equation}
  \label{eq:g1d}
  g_{\rm 1D} = \frac{2}{3} \lkl \frac{\omega_y}{\omega_z} \rkl ^{\frac{3}{5}}
  \lkl \frac{15 N a_0}{a_z} \rkl ^{\frac{3}{5}} |m_{\rm trap}|^{\frac{2}{5}}
\end{equation}
with $a_z = \sqrt{\hbar/(M\omega_z)}$.
The 1D version of Eq. (\ref{eq:gpe3d}) now reads
\begin{eqnarray}
  \label{eq:gpe1d}
  i \hbar \frac{\partial}{\partial t} \psi_m(z, t) &=&
  \lkl -\frac{\hbar^2}{2M}\frac{\partial^2}{\partial z^2}
       + V_{m,{\rm eff}}(z,t)\rkl \psi_m(z, t) \nonumber \\
  && \hspace{-1.3cm}+ \hbar \Omega
  \sum_{m'} \lkl c_{m'} \delta_{m,m'+1} + c_m \delta_{m,m'-1} \rkl
  \psi_{m'} (z, t).
\end{eqnarray}

\begin{figure}[htbp]
%  \begin{center}
    \epsfxsize=0.5\textwidth
    \epsffile{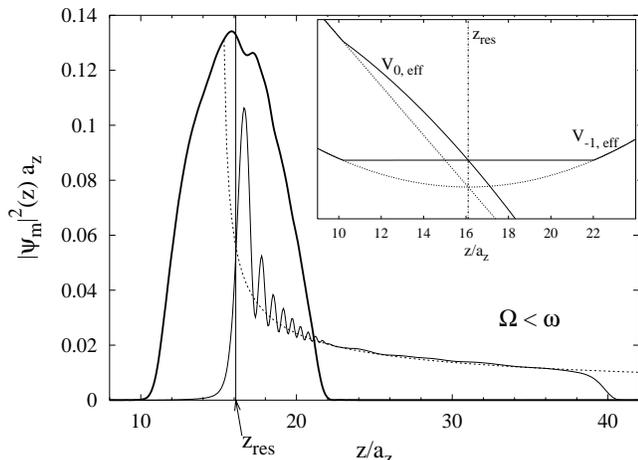}
    \caption{Plot of the densities for $m=-1,0$ ($F=1$) in the trap
      for weak outcoupling. The trapped state (thick line, scaled down
      by a factor 20) is only disturbed slightly, whereas the
      untrapped state (thin line) can be approximated by an Airy
      function (dotted line shows the magnitude).  The inset shows the
      effective potentials.}
    \label{fig:effpots}
%  \end{center}
\end{figure}

Gravitation leads to a {\em sag} of the trapped atomic cloud. The
shift of the minimum of the potentials away from the trap center at
$z=0$ depends on $m$ and is given by $z_{m,{\rm sag}} =
g/(|m|\omega_z^2)$. The effective potentials always cross at $z_{\rm
  res} = \pm a_z \sqrt{2\Delta / (\hbar\omega_z)}$. Usually, only the
positive sign plays a role because the negative value is outside the
condensate (cf. inset in Fig.~(\ref{fig:effpots})).

For weak coupling $\Omega < \omega_z$ the trapped atoms are only
disturbed slightly by the outcoupling process \cite{STEC98:1}, after a
certain propagation time the wavefunctions are quasi-stationary. In
Fig.~ \ref{fig:effpots} we plot the densities of the $m=-1,0$ states
for $F=1$ in such a situation. The outcoupling takes place primarily
at the resonance point $z_{\rm res}$. The untrapped state can be
approximated by an Airy function, its strength being determined by the
outcoupling rate which is known analytically \cite{STEC98:1,STEC97:1}.

Why is the location of maximal outcoupling identical to $z_{\rm res}$?
This is not so obvious; if one considers the same situation without
the mean-field contribution in $V_{m,{\rm eff}}$, the density maximum
of the untrapped state is located at the classical turning point in
its potential, which depends on the energy (analogous to
photo-dissociation of molecules). If the mean-field
contribution is present like in Eq.~(\ref{eq:gpe1d}) the classical
turning point is shifted to the resonance point
$z_{\rm res}$ (see inset in Fig. (\ref{fig:effpots})), and hence the
point of maximum density coincides with $z_{\rm res}$.

\section{Numerical results}
\label{sec:results}

In this section we present some numerical results obtained by
propagating the set of coupled GPEs Eq.~(\ref{eq:gpe1d}) and compare them to
experimental data for the same set of parameters. The time propagation
was accomplished by a usual split-operator method using FFT on a
1D~grid. The starting condition is always the condensate ground state
obtained by imaginary time propagation.

The basic parameters in our calculations are taken from
\cite{BLOCH99:1}, the $|F=2,\,m=2\rangle$-state is trapped with
frequencies $\omega_{\bot} = 2\pi \times 180\,\mbox{Hz}$ and
$\omega_{||} = 2\pi \times 19\,\mbox{Hz}$.  Throughout the paper we
use harmonic oszillator units with respect to the $|m|=1$ trapping
frequency for the vertical motion $\omegatrap = 2\pi \times 127\,
\mbox{Hz}$ which happens to be the same for both $F=1$ and $F=2$
($g_{F=1} = -1/2,\, g_{F=2} = 1/2$). Thus, the length unit is $a_z =
0.95\,\mu\mbox{m}$, time is measured in $1/\omegatrap =
1.3\,\mbox{ms}$.

\begin{figure}[htbp]
    \epsfxsize=0.5\textwidth
    \epsffile{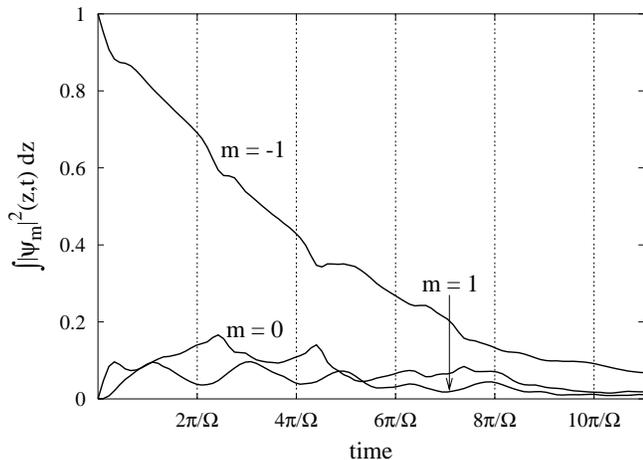}
    \caption{Time evolution of the populations for $F=1$ atom
      laser. The condensate in the $m=-1$-state is depleted very fast
      ($\Omega/\omegatrap \approx 3.5$), the time axis extends over
      $12.5\,$ms. The period of the small oscillations corresponds to
      the Rabi frequency $\Omega$.}
    \label{fig:timeevol}
\end{figure}

The Rabi frequencies used in \cite{BLOCH99:1} are all bigger than
$\omegatrap$, so we are not in a weak coupling regime. During the
outcoupling process, the condensate is depleted significantly. In
Fig.~(\ref{fig:timeevol}), one can see how fast this
takes place. The populations of the states show some small
oscillations that are reminiscent of the Rabi oscillations occuring
at very strong RF fields where there is hardly no outcoupling
\cite{BALL97,GRAH99:1}. Such oscillations can be seen in experiment as
a sort of pulses \cite{BLOCH99:privcom}.

\begin{figure}[htbp]
    \epsfxsize=0.5\textwidth
    \epsffile{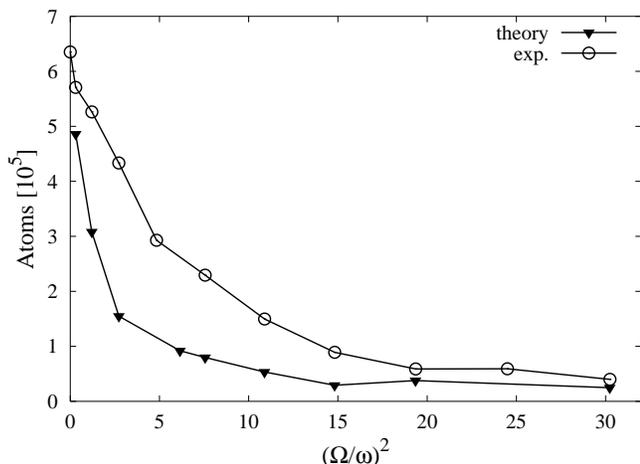}
    \caption{Number of atoms remaining in $F=2, m = 2$ after 20 ms
      of outcoupling. Initially there are $N=7 \times 10^5$ atoms. The
      detuning $\Delta$ is tuned to maximal outcoupling, so $z_{\rm res}
      \approx z_{2,{\rm sag} }$. $B_{\rm rf}$ is varied from 0.1 to
        1.0~mG leading to Rabi frequencies from 0.44 to 4.4 kHz.
        Lines are drawn for guiding the eye only.}
    \label{fig:f2Omega}
\end{figure}

\begin{figure}[htbp]
    \epsfxsize=0.5\textwidth
    \epsffile{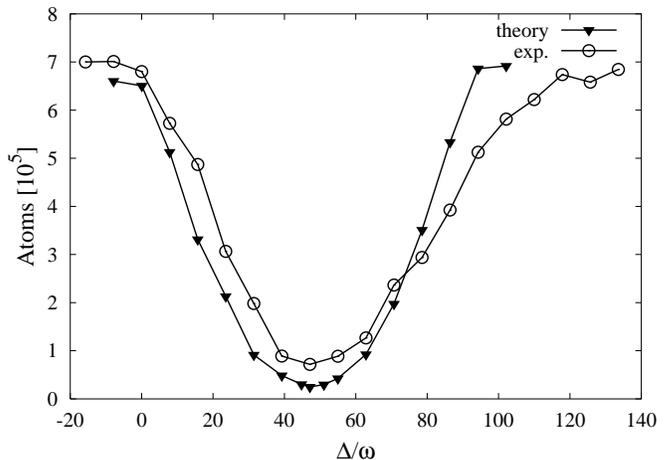}
    \caption{Same as in Fig.~(\protect\ref{fig:f2Omega}), but now
      $\Omega = 2\pi \times 700\,\mbox{Hz} = 5.5\,\omegatrap$ is kept
      fixed and the detuning $\Delta$ (resp.  radio frequency) is
      varied. The experimental data is shifted in frequency to match
      the minimum of the theoretical curve; this is necessary because
      the offset field $B_{\rm off}$ of the trap is not known with
      high enough precision. Again, lines are there for guiding the
      eye.}
    \label{fig:f2omegaspec}
\end{figure}

In \cite{BLOCH99:1} Bloch et.~al. presented measurements of the number
of particles remaining in the $|F=2,\, m=2\rangle$-state after a fixed time of
outcoupling depending both on the strength and the frequency of the
radio freqency. Figures~(\ref{fig:f2Omega}) and (\ref{fig:f2omegaspec})
display our numerical data for their parameters and
the experimental values. In the calculations we took only the
$m=2,1,0$-states into account. Tests with more states show that this
is enough, as the missing $m=-1,-2$-states are populated only very weakly.

The theoretical values fit the experimental data at least
qualitatively. The differences are most probably due to dimensional
effects: in the 1D model there is only a resonance point at
$z_{\rm res}$, whereas in 3D the outcoupling takes place on a
resonance surface defined by the distance
$r_{\rm res}=z_{\rm res}$ to the trap center. Accordingly, the
outcoupling rate in 3D depends on the density of the
trapped state averaged over the resonance surface which is smaller
than the density on the $z$-axis. Thus, the outcoupling yield in 3D is 
smaller than in 1D because there is only the ``on-axis'' density.

\begin{figure}[htbp]
    \epsfxsize=0.5\textwidth
    \epsffile{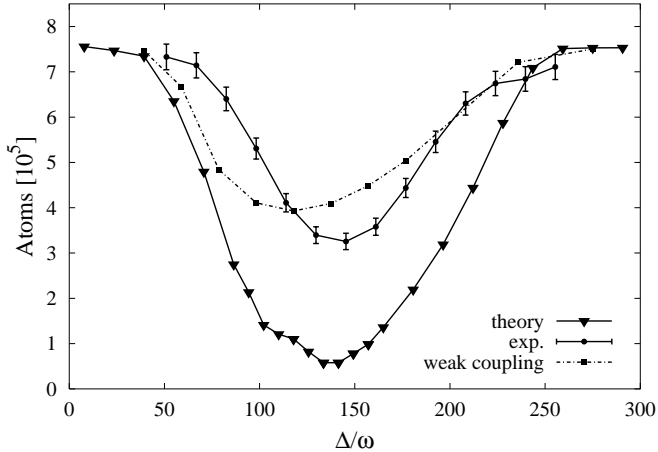}
    \caption{Same as in Fig.~(\protect\ref{fig:f2omegaspec}) but for
      $F=1, m = -1$. and $\Omega = 2\pi \times 312\,\mbox{Hz} =
      2.4\,\omegatrap$. The curve at weaker coupling ($\Omega =
      1.1\,\omegatrap$) clearly exhibits an asymmetry reflecting the
      density distribution of the trapped state (see text). Mind also the
      bump around $\Delta = 120\,\omegatrap$ in the ``theory'' curve.}
    \label{fig:f1omegaspec}
\end{figure}

This effect is even stronger for the $F=1$ results shown in
Fig.~(\ref{fig:f1omegaspec}); the experimental data are again from the
H\"ansch group \cite{BLOCH99:privcom}. In this case, all three Zeeman
sublevels were propagated because they are all significantly populated.
 (see e.g. Fig.~(\ref{fig:timeevol})). We have tried to
quantify the aforementioned difference between the 1D- and
3D-situation including gravity at least in the weak coupling limit in
the spirit of the perturbational rates from
\cite{STEC98:1,STEC97:1} but have not found a conclusive answer.

A closer look on the spectral data in Figs.~(\ref{fig:f2omegaspec})
and (\ref{fig:f1omegaspec}) reveals two further details: from a weak
coupling analysis one expects the curves in these figures to reflect
directly the density distribution of the trapped states transfered to
the frequency domain via $z =
a_z\sqrt{2\Delta/(\hbar\omegatrap)}$. This leads to an asymmetry
around the minimum that can be seen in the curve at weaker coupling
shown in Fig.~(\ref{fig:f1omegaspec}).

The second detail concerns the small bump visible in the
``theory''-curve in Fig.~(\ref{fig:f1omegaspec}) around $\Delta = 120\,
\omegatrap$. It occurs only for intermediate couplings. We have no
explanation for this feature.

\section{Conclusions and outlook}
\label{sec:conclusion}

In this paper, we have described a numerical analysis for an RF-atom
laser. The output yield of such a device can be determined in a
satisfactory way though the restriction to one dimension does not
allow for a exact quantitative comparison with the real world data
form experiment.

What is lacking is a more analytical treatment of the outcoupling
process for intermediate coupling strengths, which would probably also
explain the bump in Fig.~(\ref{fig:f1omegaspec}). Furthermore, it would be
nice to have a at least 2D picture for the outcoupled wavefunction to
discuss things like the beam divergence. A 2D treatment would also
be able to account for the long axis of the trap where collective
excitations of the condensate can occur much easier than in the other
two directions (cf. \cite{JAPHA99:1}).

Another intersting subject is an atom laser driven by two radio
frequencies. This may lead to a pulsed operation mode like in a
mode-locked laser. We are investigating such setups, results will be
published elsewhere.

\acknowledgements
  We thank Immanuel Bloch and Tilmann Ess\-linger for fruitful
  discussions and providing us with their data. The support of
  Deutsche Forschungsgmeinschaft under contract Sche 128/7-1 is
  gratefully acknowledged.

%\bibliographystyle{appphysb}
%\bibliography{lit}

\end{document}